\documentclass[aps,prb,twocolumn,superscriptaddress]{revtex4-2}
\usepackage{graphicx}
\usepackage{physics}
\usepackage[version=4]{mhchem}
\usepackage[american]{babel}
\usepackage{color}
\usepackage{tabularx}
\usepackage{multirow}
\usepackage{amsfonts}
\usepackage{amssymb}
\usepackage{ulem}
\usepackage[]{amsmath}
\usepackage[dvipsnames]{xcolor}	
\usepackage{natbib}
\usepackage{braket} 

\usepackage[bookmarks=true,															
bookmarksnumbered=true,										
colorlinks=true,
urlcolor=black,
linkcolor=black,
citecolor=blue]{hyperref}		    	
\newcommand*{\sect}{Sec.\,}

\newcommand*{\fig}{Fig.\,}
\newcommand*{\eq}[1]{Eq.\,(#1)}
\newcommand*{\tab}{Tab.\,}
\newcommand*{\reftaken}{Ref.\,}

\newcommand*{\ie}{\textit{i.e.\,}}
\renewcommand{\vec}[1]{\boldsymbol{#1}}

\newcommand*{\degree}{$^\circ$}

\newcommand{\JCNSMLZ}{J\"{u}lich Centre for Neutron Science (JCNS) at Heinz Maier-Leibnitz Zentrum (MLZ), Forschungszentrum J\"{u}lich GmbH, 85747 Garching}
\newcommand{\TUM}{Physik-Department, Technische Universit\"{a}t M\"{u}nchen (TUM), 85748 Garching, Germany}
\newcommand{\CUni}{Charles University, Faculty of Mathematics and Physics, Department of Condensed Matter Physics, 121 16, Praha, Czech Republic}
\newcommand{\MLZTUM}{TUM at MLZ, 85748 Garching, Germany}
\newcommand{\Aachen}{Institute of Crystallography, RWTH Aachen University, 52056 Aachen, Germany}
\newcommand{\TUO}{IT4Innovations, V\v{S}B-Technical University of Ostrava, 708 00 Ostrava, Czech Republic}
\newcommand{\ILL}{Institut Laue-Langevin, 38042 Grenoble, France}
\newcommand{\ISIS}{ISIS Neutron and Muon Source, STFC Rutherford Appleton Laboratory, Harwell Campus, Didcot, Oxon OX11 0QX, United Kingdom}
\newcommand{\ZQE}{Center for QuantumEngineering, TUM, 85748 Garching, Germany}
\newcommand{\MCQST}{Munich Center for Quantum Science and Technology, TUM, 85748 Garching, Germany}
\newcommand{\STUni}{School of Physical Science and Technology, ShanghaiTech University, Shanghai 201210, China}
\newcommand{\STech}{ShanghaiTech Laboratory for Topological Physics, ShanghaiTech University,Shanghai 201210, China}

\begin{document}

\title{Orthorhombic deformation enables long-range magnetic order in CePdAl$_3$}

\author{M.~Stekiel}
\email{m.stekiel@fz-juelich.de}
\affiliation{\JCNSMLZ}
\affiliation{\TUM}

\author{P.~\v{C}erm\'{a}k}
\affiliation{\JCNSMLZ}
\affiliation{\CUni}

\author{C.~Franz}
\affiliation{\JCNSMLZ}
\affiliation{\MLZTUM}

\author{M.~Meven}
\affiliation{\JCNSMLZ}
\affiliation{\Aachen}

\author{D.~Legut}
\affiliation{\CUni}
\affiliation{\TUO}

\author{W.~Simeth}
\affiliation{\TUM}

\author{U.B.~Hansen}
\affiliation{\ILL}

\author{B.~F{\aa}k}
\affiliation{\ILL}

\author{S.~Weber}
\affiliation{\TUM}

\author{R.~Sch\"{o}nmann}
\affiliation{\TUM}

\author{V.~Kumar}
\affiliation{\TUM}

\author{K.~Nemkovski}
\affiliation{\JCNSMLZ}
\affiliation{\ISIS}

\author{H.~Deng}
\thanks{Present address: \STUni, \STech}
\affiliation{\JCNSMLZ}
\affiliation{\Aachen}

\author{A.~Bauer}
\affiliation{\TUM}
\affiliation{\ZQE}

\author{C.~Pfleiderer}
\affiliation{\TUM}
\affiliation{\MLZTUM}
\affiliation{\ZQE}
\affiliation{\MCQST}

\author{A.~Schneidewind}
\affiliation{\JCNSMLZ}


\date{\today}

\begin{abstract}
	We investigate the effect of structural deformation on the magnetic properties of orthorhombic CePdAl$_3$ in relation to its tetragonal polymorph. Utilizing x-ray and neutron diffraction we establish that the crystal structure has the $Cmcm$ space group symmetry and exhibits pseudo-tetragonal twinning. According to density-functional calculations the tetragonal-orthorhombic deformation mechanism has its grounds in relatively small free enthalpy difference between the polymorphs, allowing either phase to be quenched and fully accounts for the twinned microstructure of the orthorhombic phase. Neutron diffraction measurements show that orthorhombic CePdAl$_3$ establishes long-range magnetic order below $T_\mathrm{N}$=5.29\,(5)\,K characterized by a collinear, antiferromagnetic arrangement of magnetic moments. Magnetic anisotropies of orthorhombic CePdAl$_3$ arise from strong spin-orbit coupling as evidenced by the crystal-field splitting of the $4f$ multiplet, fully characterised with neutron spectroscopy. We discuss the potential mechanism of frustration posed by antiferromagnetic interactions between nearest neighbours in the tetragonal phase, which hinders the formation of long-range magnetic order in tetragonal CePdAl$_3$. We propose that orthorhombic deformation releases the frustration and allows for long-range magnetic order.
\end{abstract}

\maketitle


\section{Introduction}

Symmetry is a fundamental characteristic of any physical system, determining the presence of certain properties and their anisotropy \cite{Curie-JdP-1894}.
For crystalline compounds, where the crystal structure reflects the symmetry of the system, determining the structure-property relations is of fundamental importance for understanding emerging phenomena and tailoring materials with desired functionalities.
In particular, the class of ternary intermetallics with the ThCr$_2$Si$_2$ parent structure comprises hundreds of members \cite{Shatruk-JSSC-2019}, among which there are systems that exhibit compelling properties such as unconventional superconductivity \cite{Steglich-PRL-1979, Bauer-PRL-2004, Takeuchi-JPCondMatt-2004, Kimura-PRL-2005,  Pfleiderer-RevModPhys-2009}, heavy-fermion states \cite{Egetenmeyer-PRL-20012, Hossain-PRB-2005, Palstra-PRL-1985}, non-Fermi-liquid behaviour \cite{Stewart-RevModPhys-2001}, and complex magnetic order \cite{Klicpera-PRB-2015, Adroja-PRB-2015, Stekiel-PRR-2023, Anand-PRB-2021, Khanh-NatureNano-2020, Takagi-NatComm-2022}.

Here, we investigate Ce$TX_3$ systems, the Ce-113 family, where $T$ is a transition metal and $X$ a $p$-block element.
We focus on the long-range magnetic order, which emerges as a combination of single-ion anisotropies and exchange interactions of cerium $4f$ electrons.
The crystal-electric field (CEF) splits the $4f$ multiplet introducing single-ion anisotropies \cite{Stevens-PPS-1952}, while exchange interactions typically emerge from indirect interaction between $f$-electrons mediated by the conduction band formalised as Rudermann-Kittel-Kasuya-Yosida (RKKY) interaction \cite{vVleck-RevModPhys-1962}.
Our scientific interest is driven by the observation of multiple types of magnetic modulations in members of the Ce-113 family sharing the same crystal structure, described with the tetragonal $I4mm$ space group.
Despite structural identity, the types of magnetic order include amplitude \cite{Hillier-PRB-2012, Anand-PRB-2018, Smidman-PRB-2013, Anand-PRB-2021, Klicpera-PRB-2015}, spatial modulations \cite{Adroja-PRB-2015, Stekiel-PRR-2023}, with the moments along the four-fold axis \cite{Hillier-PRB-2012, Anand-PRB-2018, Smidman-PRB-2013}, or perpendicular to it \cite{Anand-PRB-2018, Adroja-PRB-2015, Klicpera-PRB-2015}.
In addition, the tetragonal CePdAl$_3$ member of the family was reported to defy long-range magnetic ordering down to 0.1\,K \cite{Franz-JAC-2016}, and presents itself as an interesting candidate to study the components required to establish and characterise long-range magnetic order in the Ce-113 family.

Previous reports on the magnetic properties of CePdAl$_3$ are based on specific heat, magnetization, magnetic susceptibility and resistivity measurements as a function of temperature and applied field \cite{Franz-JAC-2016, Schank-JAC-1994, Kumar-PRRes-2023}.
They conclude that the tetragonal variant of CePdAl$_3$ with the $I4mm$ space group does not exhibit long-range magnetic order down to 0.1\,K \cite{Franz-JAC-2016, Schank-JAC-1994}.
However, the orthorhombic variant of CePdAl$_3$, obtained by a modified synthesis protocol, shows features characteristic of long-range magnetic ordering at low-temperature \cite{Schank-JAC-1994, Kumar-PRRes-2023}.
In particular, the specific heat exhibits a broad transition at $T_\mathrm{N_1}$=5.6\,K and a lambda anomaly at $T_\mathrm{N_2}$=5.4\,K.
Magnetic properties of orthorhombic CePdAl$_3$ are characteristic of an easy-axis antiferromagnet (AF), in which the ground state multiplet of $4f$ electron is split into three doublet states \cite{Kumar-PRRes-2023}.

In this study we report on the crystal and magnetic structure of orthorhombic CePdAl$_3$ as well as the characterization of the $4f$ multiplet split by the crystal electric field.
The samples studied here originate from the same synthesis batch as in \reftaken\cite{Kumar-PRRes-2023}, where the system was characterized by means of specific heat, magnetic susceptibility, and magnetization under applied field down to 2\,K.
Parts of this study, specifically the twinning scheme and lattice type drawn from x-ray diffraction measurements were reported in \reftaken\cite{Kumar-PRRes-2023}.
Here, we elaborate further on these aspects, together with a precise report on the space group assignment, as well as the crystal structure solution and refinement.
Based on density functional theory (DFT) calculations we investigate the structural stability of the tetragonal and orthorhombic polymorphs of CePdAl$_3$.
We report a neutron diffraction study at low temperature and identify a long-range magnetic order with a collinear, antiferromagnetic arrangement of magnetic moments.
Based on inelastic neutron scattering measurements, we determine the CEF scheme utilizing tools specifically developed to analyse the measured data.
This approach to data analysis allowed us to constrain the parameters describing the crystal-electric field in orthorhombic CePdAl$_3$ and determine a unique set of parameters.
The characteristics of the orthorhombic CePdAl$_3$ determined by various techniques in this and the related study \cite{Kumar-PRRes-2023} provide a consistent picture of the structural and magnetic properties of the system.


\section{Methods}

\subsection{Sample synthesis and characterization}

Orthorhombic CePdAl$_3$ was grown in an optical furnace utilizing the floating-zone technique \cite{Neubauer-RSI-2011, Bauer-RSI-2016}.
The synthesis protocol is described in \reftaken\cite{Kumar-PRRes-2023}.
Crystals from the same rod were used to perform measurements described in \reftaken\cite{Kumar-PRRes-2023}, where they were thoroughly characterized by means of specific heat, magnetization, and ac susceptibility measurements.

According to the experimental technique, samples of specific size were extracted from the crystalline rod.
X-ray diffraction measurements were performed on six crystals with cuboid shape and dimensions varying between 10\,$\mu$m and 50\,$\mu$m.
Neutron diffraction measurements were conducted on a cuboid-shaped crystal with dimensions 1$\times$2$\times$4\,mm$^3$.
Neutron spectroscopy measurements were performed with a cylinder shaped crystal with diameter of 6\,mm and length 15\,mm.

\subsection{X-ray and neutron scattering}
\label{sec:methods}

X-ray diffraction measurements were performed on a Rigaku XtaLAB Synergy-S diffractometer, using a Mo x-ray source providing a monochromatic beam with a wavelength of 0.71\,\AA\ and a two-dimensional HyPix-Arc 150$^{\circ}$ detector.
The observed reflections were indexed and integrated using the data reduction program \textsc{CrysalisPro} \cite{Crysalis}.

The modulation vector of the magnetic order in CePdAl$_3$ was determined from measurements at the Diffuse Neutron scattering Spectrometer (DNS) \cite{DNS-LSF, Schweika-PhysicaB-2001} at Heinz Maier-Leibnitz Zentrum (MLZ) using polarization analysis.
The incoming neutrons wavelength was 4.2\,\AA\, covering the momentum transfer range $Q$ of 0.25\,\AA$^{-1}$$<$$Q$$<$2.7\,\AA$^{-1}$. 
The sample was rotated in 1$^\circ$ steps in a 140$^\circ$ range to map the neutron scattering intensity.
The incoming neutrons were polarized along the momentum transfer vector, which allows to distinguish between magnetic and nuclear scattering contributions on the spin-flip (SF) and non-spin-flip (NSF) channels, respectively \cite{Schweika-PhysicaB-2001, Boothroyd-2020}. 
The flipping ratio for our measurements is FR=18, which corresponds to the polarization efficiency $\epsilon$=95\%.
The data were analysed, reduced and corrected for flipping ratio using the Mantid package \cite{Mantid}.

Intensities of the nuclear and magnetic reflections required to determine the type of long-range magnetic order in CePdAl$_3$ were measured at the neutron diffractometer HEiDi \cite{HEiDi-LSF} at MLZ.
The crystal was mounted on an Eulerian cradle, the wavelength of incoming neutrons was 0.793\,\AA.
Scattered neutrons were measured by a point detector moving in the horizontal plane.
First, the sample was measured at ambient conditions, and then it was mounted in a closed cycle cryostat.
A series of measurements was performed at various temperatures down to 2.4\,K.
Intensities of the observed reflections were refined with the Jana program \cite{Jana-2014}.

The excitation spectrum of the $4f$-electron split in the crystal field of orthorhombic CePdAl$_3$ was determined from measurements \cite{data-PANTHER} at the thermal neutron time-of-flight spectrometer PANTHER \cite{Panther-ILL} at the Institut Laue Langevin.
The instrument operates with a single-crystal monochromator and a Fermi chopper which provide a monochromatic beam of incoming neutrons at required energies.
An array of $^3$He tubes forms a 2D position-sensitive detector covering 141\degree\ in-plane and 43\degree\ out of plane scattering angles.

\subsection{Density functional theory calculations}

The calculations were performed within the density functional method implemented in Vienna Ab Initio Simulation Package \cite{VASP}. 
We employed projector augmented-waves pseudopotentials utilizing the generalized gradient approximation exchange-correlation functional with the Perdew, Burke and Ernzerhof parametrization \cite{PBE}.

The electronic valence configurations for Ce, Pd, and Al, were $6s^25d^14f^1$, $5s^14d^9$, and $3s^22p^1$, respectively. 
The Brillouin zone was sampled with 245 and 220 k-points for the orthorhombic $Cmcm$ and tetragonal $I4mm$ phases, respectively, and a plane-wave energy cut-off of 480\,eV was used for both.

Structures were fully optimized, including lattice parameters and atomic positions, for selected volumes.
Convergence criteria for total energy and forces acting on each atom were set at less than 10$^{-7}$\,eV and 10$^{-5}$\,eV/\AA, respectively. 

The dynamical properties of the orthorhombic and tetragonal lattices were investigated using the quasi-harmonic approximation for lattice vibrations on a 2$\times$2$\times$2 supercell, employing Phonopy \cite{Phonopy}.
This allowed determination of the phonon density of states and subsequently the thermodynamic properties, such as the Gibbs free energy, including the electronic contribution as a function of temperature.
The transition temperature was determined similarly to recent work, see \reftaken\cite{Legut-MatChemPhys-2014} and references therein for details.

\subsection{Crystal-electric field transitions}

The crystal-electric field lifts the degeneracy of the $4f$ multiplet and splits the energy levels on scales typically accessible with neutron scattering.
The good quantum number describing this problem is the total angular momentum $J$, and the analysis follows from considering the CEF Hamiltonian
\begin{equation}
	\label{eq:hamiltonian}
	\hat{H} = \sum_{n,m} B_n^m \hat{O}_n^m.
\end{equation}
Here, $\hat{O}_n^m$ are Stevens operators, and $B_n^m$ crystal field parameters.
Stevens operators consist of total momentum operators $\hat{J_x}$, $\hat{J_y}$, $\hat{J_z}$ grouped, raised to the power $n$, and composed in a way that reflects the point symmetry of the magnetic ion \cite{Stevens-PPS-1952}.
Additionally, symmetry restrictions can constrain the values of some $B_n^m$ parameters to zero.
The Hamiltonian from \eq{\ref{eq:hamiltonian}} represented in the $\ket{m_J}$ basis, where $m_J = J, J-1,\ldots, -J$, is diagonalized, and the eigenstates $\ket{\lambda}$ are linear combinations of the $\ket{m_J}$ states.

The cross section of the neutron scattering process corresponding to the CEF transition from state $\ket{\lambda_i}$ to $\ket{\lambda_f}$ for neutron momentum transfer $\vec{Q}$ and energy transfer $E$ follows from the expectation values of the perpendicular projection of the total momentum operator $\vec{\hat{J}} = (\hat{J}_x, \hat{J}_y, \hat{J}_z)$ on the momentum transfer \cite{Boothroyd-2020}.
This is expressed as
\begin{align}
\frac{d \sigma}{d \Omega} &\propto | \braket{ \lambda_f | \vec{\hat{J}}_{\perp Q} | \lambda_i }|^2 = \nonumber \\
&= 
\sum_\alpha \left[ 1-\left(Q_\alpha/|\vec{Q}|\right)^2 \right] \left| \braket{ \lambda_f | \hat{J}_\alpha | \lambda_i } \right|^2, \label{eq:Jperp-SC}
\end{align}
where $\alpha$ indexes the Cartesian components, $\alpha =x, y, z$.

Expression (\ref{eq:Jperp-SC}) can be conveniently evaluated for polycrystalline samples, where the momentum transfer $\vec{Q}$ is angularly averaged, yielding $\frac{2}{3}\sum_\alpha \braket{ \lambda_f | \hat{J}_\alpha | \lambda_i }|^2$.
However, this approach would result in a loss of information, providing fewer constraints on determining the crystal-field parameters.

To address this limitation, we have developed tools to directly evaluate \eq{\ref{eq:Jperp-SC}} for any measured $Q-E$ point.
This allows us to qualitatively analyse neutron scattering intensity due to CEF transitions on a single-crystal sample and impose stronger constraints on the $B_n^m$ parameters.
The code used in the analysis is available under the public repository \textsc{crysfipy} \cite{crysfipy}, developed as an extension of the previous version of the project \cite{crysfipy-Petr}.

For the case of a cerium atom with $J$=$\frac{5}{2}$ that exhibits $m2m$ ($C_{2v}$) point symmetry, there are five non-zero $B_n^m$ parameters.
At the same time, $J$=$\frac{5}{2}$ signifies a Kramers' ion, and in this case, the CEF energy level scheme consists of three doublet states.
The standard method of analysing the transition energies and transition intensity ratio on polycrystalline sample would yield only three constraints, while our method provides additional ones.


\section{Experimental results}

\subsection{Pseudo-tetragonal twinning}
\label{sec:twinning}

Here, we elaborate on details of our recent study \cite{Kumar-PRRes-2023} and present an extended analysis of the twinning and crystal structure determination

\begin{figure}[t]
	\begin{center}
		\includegraphics[width=0.48\textwidth]{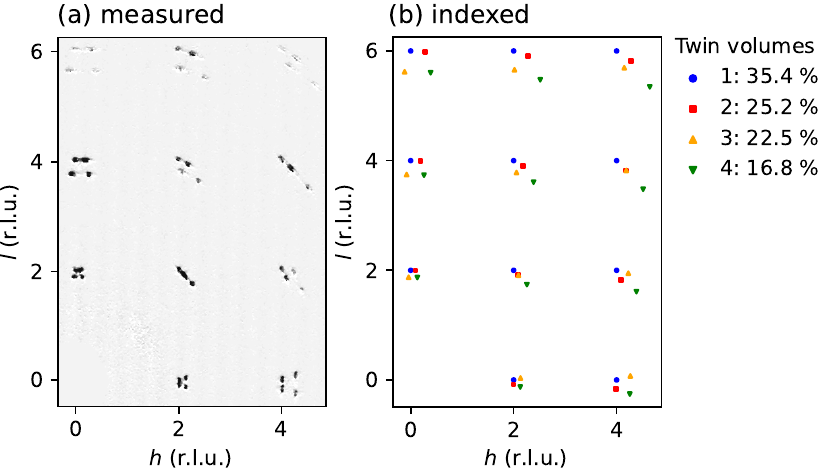}
	\end{center}
	\caption{Diffractogram of orthorhombic CePdAl$_3$. (a) Measured x-ray scattering intensity reconstructed on the ($h0l$) plane in reciprocal space. (b) Indexing of reflections from (a) with four identical, orthorhombic lattices grouped by colour. Indices follow the majority twin (blue circles). Characteristic splitting of reflections reveals pseudo-tetragonal twinning.}
	\label{fig:twinning-reciprocal}
\end{figure}

We investigated the x-ray diffraction pattern of six micrometer-sized crystals of CePdAl$_3$ at ambient conditions.
All patterns consist of a characteristic set of split reflections as shown in \fig\ref{fig:twinning-reciprocal}.
By indexing measured reflections, we disentangle split and overlapped spots into four twins of the same orthorhombic lattice with parameters $a=6.3785$\,(6)\,\AA, $b=10.402$\,(2)\,\AA, and $c=5.972$\,(2)\,\AA.
Based on reflection conditions, the diffraction symbol is $C \text{-} c \text{-}$, which, following crystallographic convention, establishes $b$ as the long axis.

\begin{figure}[t]
	\begin{center}
		\includegraphics[width=0.48\textwidth]{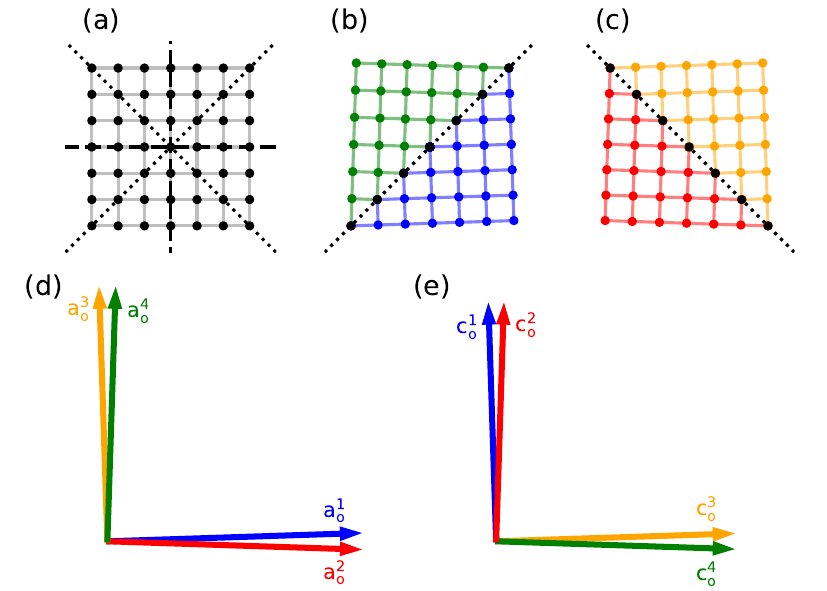}
	\end{center}
	\caption{Twinning scheme in orthorhombic CePdAl$_3$. (a) Undistorted lattice. Diagonal mirror planes are indicated with dotted lines, normal mirror planes with dashed lines. (b) Twins related by one diagonal mirror plane and (c) the other diagonal mirror plane. (d) Orientation of the $(1\,0\,0)$ and (e) the $(0\,0\,1)$ lattice vectors among twins. The scale of the distortion in the figures is the same as determined from the single-crystal x-ray diffraction measurements.}
	\label{fig:twinning-real}
\end{figure}

The characteristic diffraction pattern with sets of three spots around the $(h0h)$ reflections and four spots around $(h0l)$ ($h \neq l$) reflections is indicative of pseudo-tetragonal twinning with diagonal mirror planes defining the twinning laws, as shown in \fig\ref{fig:twinning-real}.
The relative orientation of all twins follows from the mismatch angle $\phi_m$, defined as the angle between close-lying, equivalent axes, such as $a_1$ and $a_2$ in \fig\ref{fig:twinning-real}(d).
Taking the twin with index 1 as the reference, twin 2 is related to twin 1 by a right-handed rotation of an angle $-\phi_m$ around the perpendicular $b$ axis, twin 3 by rotation of 90\degree, and twin 4 by rotation of $90-\phi_m$.
The mismatch angle determined from the orientation matrices is $\phi_m = 3.7 (3)$\degree.

Despite pseudo-tetragonal symmetry, the twin volumes are not equal.
Instead, in the crystal selected for structure determination, twins 1 and 2 represent 35.4\,\% and 25.2\,\% of the crystal volume, respectively, while twins 3 and 4 represent 22.5\,\% and 16.8\,\%, respectively.
Other crystals had similar twin volumes, with differences less than 4\,\% for any twin component.
This has important consequences for the anisotropic properties, which was taken into account in \reftaken\cite{Kumar-PRRes-2023}.

\subsection{Crystal structure}
\label{sec:cryst_struct}

Extinction conditions are characteristic of the symbol $C \text{-} c \text{-}$, for which possible space groups are the centrosymmetric $Cmcm$ (number 63), as well as non-centrosymmetric $Cmc2_1$ (36), and $C2cm$ (40).
A significant amount of overlapping reflections does not allow for a reliable determination of the Flack parameter and directly verify the presence of inversion symmetry.

\begin{figure}[t]
	\begin{center}
		\includegraphics[width=0.48\textwidth]{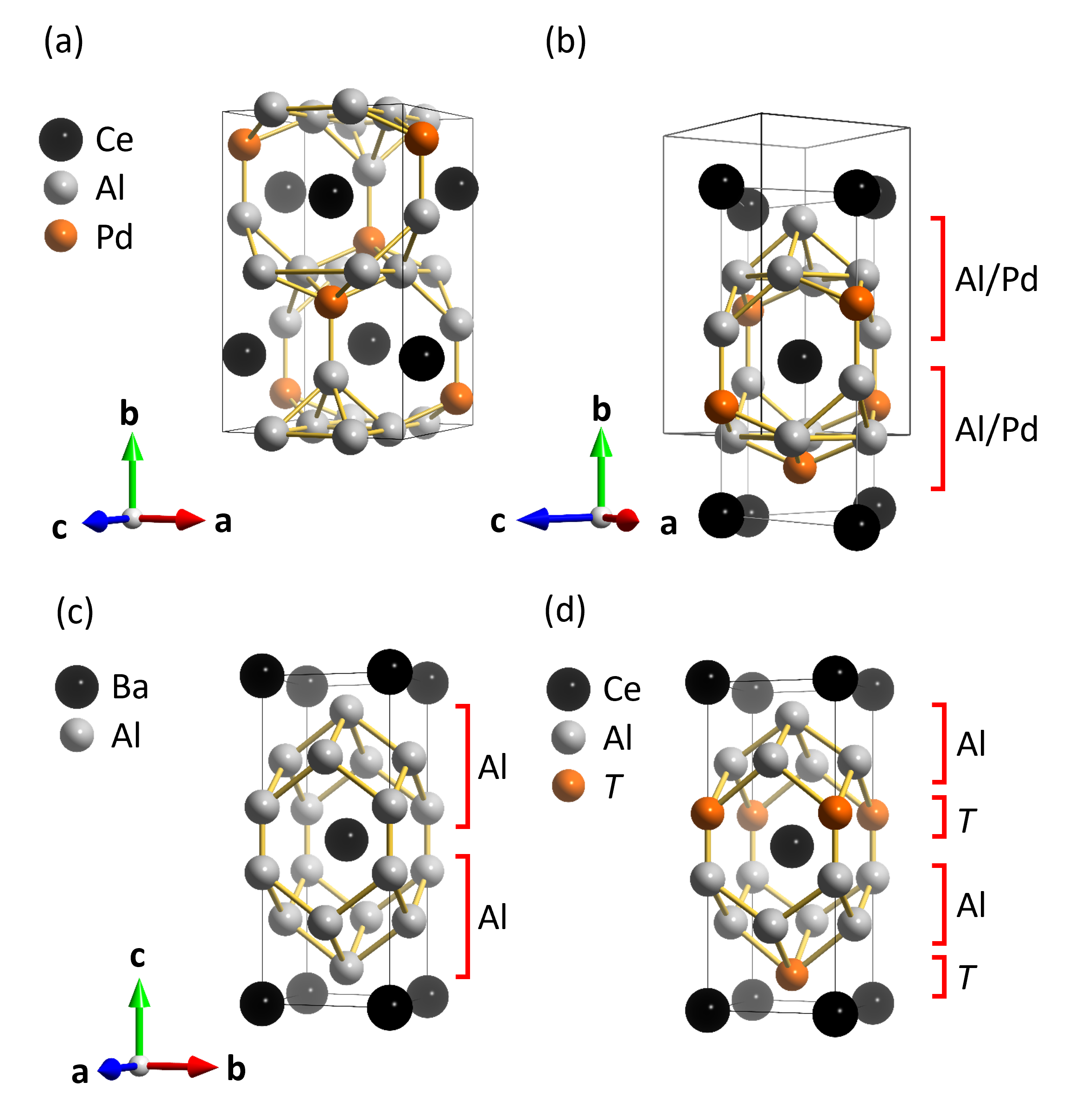}
	\end{center}
	\caption{Crystal structure of orthorhombic CePdAl$_3$ and related compounds. $Cmcm$ structure of CePdAl$_3$ with atoms in the (a) orthorhombic unit cell  and (b) projected to the tetragonal unit cell. (c) Centrosymmetric, tetragonal BaAl$_4$ structure with space group $I4/mmm$. (d) Non-centrosymmetric, tetragonal Ce$T$Al$_3$ structure with $I4mm$ space group. Note the increasing level of distortion in the neighbourhood of the cerium atom in the $I4/mmm \rightarrow I4mm \rightarrow Cmcm$ series. Red brackets enclose atoms forming uni-atomic layers in (c) and (d), and mixed Al-Pd layer in CePdAl$_3$. The unit cell is marked with a solid black line, close-lying atoms are connected with yellow lines.}
	\label{fig:crystal-structure}
\end{figure}

Crystal structure determination of CePdAl$_3$ by direct methods results in the PbSbO$_2$Cl structure type, reported also for the related CeAgAl$_3$ \cite{Franz-JAC-2016}. The atomic arrangement of the orthorhombic CePdAl$_3$ is shown in \fig\ref{fig:crystal-structure}, with detailed parameters in \tab\ref{tab:crystal-structure}.

Refinement of measured intensities within the space group $Cmcm$ corroborates the structure solution by low $R$-factors, of $R_{int}$=4.92 for the data quality, and $R_\mathrm{obs}$=3.90 for the refinement quality.
Following the observation of atomic disorder in tetragonal Ce$T$Al$_3$ systems \cite{Franz-JAC-2016, Stekiel-PRR-2023}, we (i) split the atomic positions of Pd or Al atoms and (ii) introduced occupational disorder between the Pd and Al sites.
Neither (i), (ii) or combination of both improved the quality of the refinement.

Another attempt to model disorder was by considering the non-centrosymmetric space groups $Cmc2_1$ and $C2cm$, as they lift constraints on certain atomic positions.
The refinement quality factors do not differ significantly, $R_\mathrm{obs}^{Cmc2_1}$=3.93 $R_{obs}^{C2cm}$=3.71, despite introducing additional parameters.
The shifts of atomic positions are smaller than 0.03\,\AA\ with respect to the $Cmcm$ model.

An independent investigation of the crystal structure was based on neutron diffraction measurements.
Large data sets were collected at 300 and 10\,K, each contains a few hundred reflections.
Refinements result in the same crystal structure as described in \tab\ref{tab:crystal-structure} with negligible differences in atomic positions, smaller than 1\% of the relative difference.

In conclusion, we describe the crystal structure of orthorhombic CePdAl$_3$ with the highest symmetry suitable space group $Cmcm$ and atomic coordinates listed in \tab\ref{tab:crystal-structure}, without evidence of atomic disorder.

The description of the crystal structure is facilitated by comparing it with the tetragonal $I4mm$ structures of Ce$T$Al$_3$, $T$=Cu, Au, Pt \cite{Franz-JAC-2016}, and the ThCr$_2$Si$_2$ aristotype represented by BaAl$_4$ with $I4/mmm$ space group \cite{Andress-ZfMetall-1935}, shown in \fig\ref{fig:crystal-structure}.
In those compounds, the crystal structure consists of a three-dimensional $X$-Al network, where $X$ is either $T$ or Al, forming cages occupied by the central Ce atom.

In the high-symmetry BaAl$_4$ structure, the cerium atoms are neighboured by aluminium atoms, as shown in \fig\ref{fig:crystal-structure}(c).
By removing the inversion symmetry, the $I4/mmm$ structure transforms into the Ce$T$Al$_3$ type, see \fig\ref{fig:crystal-structure}(d), adapted by tetragonal CePdAl$_3$.
In this configuration, cerium atoms are surrounded on one side by $T$ atoms, and on the other by aluminium atoms, forming uni-atomic layers, highlighted in \fig\ref{fig:crystal-structure}.

Alternatively, orthorhombic distortion of the BaAl$_4$ structure and further splitting of atomic position results in the $Cmcm$ structure, adapted by orthorhombic CePdAl$_3$.
The inversion symmetry is preserved, but with respect to the corners of the orthorhombic unit cell, where no atom resides.
As a result, the environment of cerium atom in the $Cmcm$ structure is more distorted than in Ce$T$Al$_3$, compare \fig\ref{fig:crystal-structure}(b) and (d), and the layers between cerium atoms comprise of mixed palladium and aluminium atoms.

\begin{table}[t]
\centering
\caption{Details of the orthorhombic CePdAl$_3$ crystal structure based on refinement of x-ray diffraction data at ambient temperature. Space group $Cmcm$,  $a=6.3785$\,(6)\,\AA, $b=10.402$\,(2)\,\AA, $c=5.972$\,(2)\,\AA. The number of measured, independent, and observed reflections is 2509, 280, and 260, respectively. The reflection merge factor $R_\mathrm{int}$=4.92, observed reflections are with $I>3\sigma(I)$. The refinement was performed on $F^2$ with $R_\mathrm{obs}$=3.90 and $wR_\mathrm{obs}$=8.47. The columns contain the name of the elements, Wyckoff site symbols, atomic coordinates, and displacement parameters, respectively. Numbers without errors are restricted by symmetry.}
\label{tab:crystal-structure}
\begin{tabularx}{0.48\textwidth}{X X X l}
	\hline\hline
	Element & Site & Position & U (\AA$^2$) \\
	\hline
	\multirow{3}{*}{Ce} & \multirow{3}{*}{4$c$} &
	x=0 & U$_{11}$=0.0313 (5) \\
	& & y=0.7647 (1) & U$_{22}$=0.0275 (5) \\
	& & z=0.25 & U$_{33}$=0.0333 (5) \\
	\hline
	\multirow{3}{*}{Pd} & \multirow{3}{*}{4$c$} &
	x=0 & U$_{11}$=0.0317 (6) \\
	& & y=0.0949 (1) & U$_{22}$=0.0285 (7) \\
	& & z=0.25 & U$_{33}$=0.0392 (7) \\
	\hline
	\multirow{3}{*}{Al} & \multirow{3}{*}{4$c$} &
	x=0 & U$_{11}$=0.0257 (19) \\
	& & y=0.3336 (5) & U$_{22}$=0.0400 (30) \\
	& & z=0.25 & U$_{33}$=0.0251 (18) \\
	\hline
	\multirow{3}{*}{Al} & \multirow{3}{*}{8$e$} &
	x=0.2795 (5) & U$_{11}$=0.0309 (15) \\
	& & y=0 & U$_{22}$=0.0289 (17) \\
	& & z=0 & U$_{33}$=0.0328 (15) \\
	& & & U$_{23}$=0.0008 (12) \\
	\hline\hline
\end{tabularx}	
\end{table}

\subsection{Structural stability of the orthorhombic phase}
\label{sec:dft}

In order to verify the thermodynamic stability of the orthorhombic CePdAl$_3$, we investigated its Gibbs free energy, $G$, as a function of temperature.
As an alternative structure we considered the tetragonal $I4mm$ variant, reported by \cite{Franz-JAC-2016} and depicted in \fig\ref{fig:crystal-structure}(d).

\begin{figure}[t]
	\begin{center}
		\includegraphics[width=0.48\textwidth]{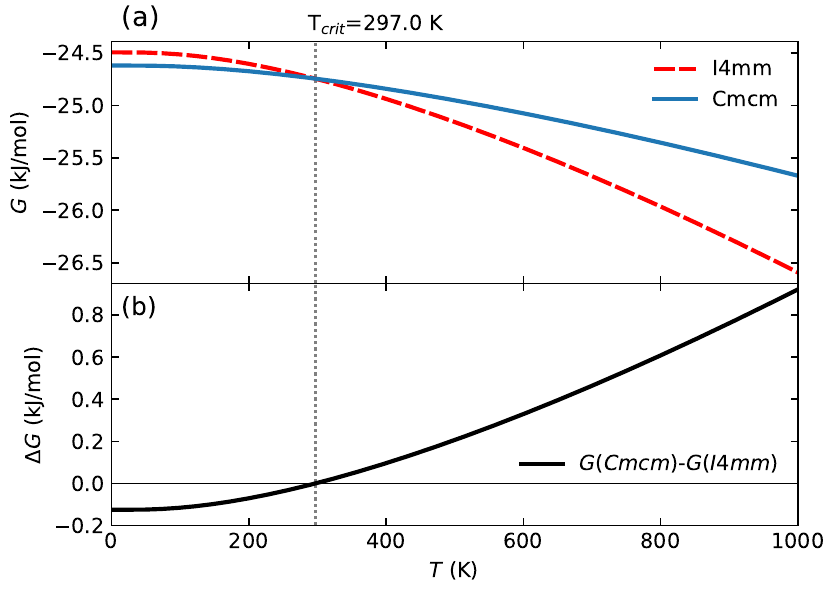}
	\end{center}
	\caption{Gibbs free energy $G$ of the structural variants of CePdAl$_3$. (a) $G$ as a function of temperature for the tetragonal $I4mm$ (dashed red line) and orthorhombic $Cmcm$ (solid blue line) variants. (b) Difference in $G$ between the orthorhombic and tetragonal variants. The dotted vertical line at $T_\mathrm{crit}$=297\,K marks the point with $\Delta G$=0, where a structural transition would occur.}
	\label{fig:therm-G}
\end{figure}

The Gibbs free energy as a function of temperatures is shown in \fig\ref{fig:therm-G}.
The phases have equal energy at $T_\mathrm{crit}$=297\,K, which marks the structural transition temperature between the $I4mm$ and $Cmcm$ structures.
Above $T_\mathrm{crit}$ the tetragonal phase has lower energy and is thus stable, while the orthorhombic phase is metastable.
Below $T_\mathrm{crit}$ the situation is reversed, and the orthorhombic variant is thermodynamically stable.

The difference in Gibbs energy, $\Delta G$, between those phases is relatively small, less than 1\,kJ/mol in the temperature range studied, see \fig\ref{fig:therm-G}(b).
Our calculations do not determine the activation energy between the considered phases, however, this should be relatively high, given the displacive character of the transition, which involves rearrangement of the Al-Pd atomic positions, as depicted in \fig\ref{fig:crystal-structure} panels (b) and (d).
These observations suggests that both phases can exist at ambient conditions, either in a stable or metastable form, consistent with previous reports \cite{Schank-JAC-1994, Franz-JAC-2016, Kumar-PRRes-2023}.

\subsection{Magnetic structure}
\label{sec:magnetic}

Orthorhombic CePdAl$_3$ shows two magnetic transitions at low temperatures, at $T_\mathrm{N_1}$=5.6\,K, and $T_\mathrm{N_2}$=5.4\,K \cite{Kumar-PRRes-2023}.
In order to determine the type of magnetic order we performed a series of neutron diffraction measurements.

\begin{figure}[t]
	\begin{center}
		\includegraphics[width=0.48\textwidth]{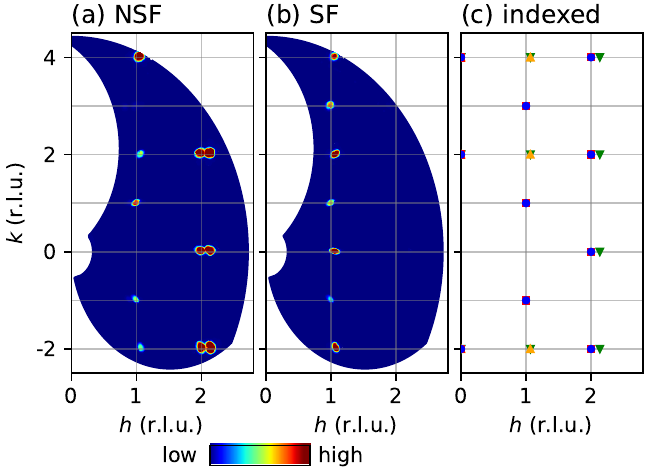}
	\end{center}
	\caption{Neutron scattering intensity in the ($hk0$) plane of CePdAl$_3$ at 3\,K. (a) Intensity of the non-spin-flip (NSF) channel, where nuclear Bragg reflections are observed, and (b) in the spin-flip channel (SF), where magnetic reflections are visible. (c) Splitting and misalignment of nuclear reflections, as inferred from the twinning. Only reflections allowed by the $Cmcm$ space group are shown in panel (c), with the indexation scheme as in \fig\ref{fig:twinning-reciprocal}.}
	\label{fig:dns}
\end{figure}

Mapping of the reciprocal space of CePdAl$_3$ by neutron scattering at 3\,K, below the magnetic transitions is shown in \fig\ref{fig:dns}.
Reflections are separated based on the nuclear or magnetic origin owing to polarization analysis, as shown in panels \fig\ref{fig:dns}(a) and \fig\ref{fig:dns}(b), respectively.

Some reflections are split or misaligned with respect to the indices of the majority twin, indicated by the grid.
The map of nuclear reflections expected for the $Cmcm$ space group within the twinning scheme described in \sect\ref{sec:twinning} is shown in \fig\ref{fig:dns}(c), and is fully consistent with observed reflections as shown in \fig\ref{fig:dns}(a).

Taking twinning into account, all magnetic reflections occur at integer positions, indicating a commensurate magnetic structure, with the magnetic unit cell of the same size as the chemical unit cell.
In particular, the magnetic reflection around the $(hk0)_1$=$(1+\delta 0 0)$ position, identified as the $(001)$ reflection from twins 3 and 4, is not allowed by the extinction rules of the $Cmcm$ space group, and is indicative of an antiferromagnetic order with magnetic moments alternating along the $(001)$ planes.

\begin{figure}[t]
	\begin{center}
		\includegraphics[width=0.49\textwidth]{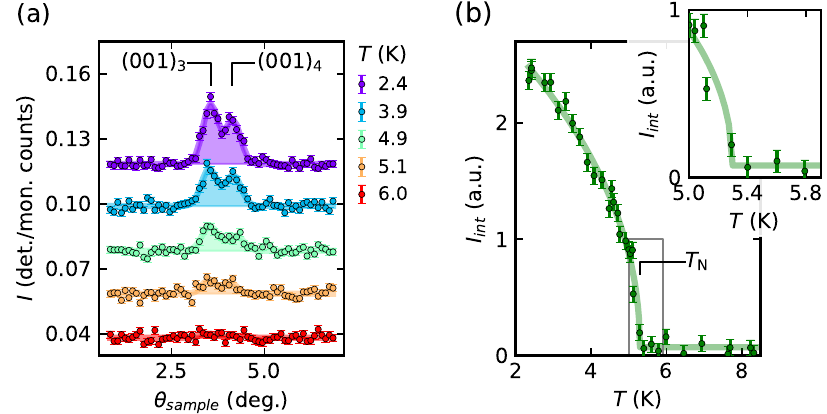}
	\end{center}
	\caption{Development of the magnetic order in CePdAl$_3$. (a) Rocking scans of the $(0\,0\,1)$ reflection at various temperatures. The splitting of the reflection is due to twinning, with the contributing twins marked. Points indicate measured data, lines are guides to the eye. (b) Integrated intensity of the $(0\,0\,1)$ reflection as a function of temperature. The fitted order parameter is shown with a solid line.}
	\label{fig:heidi}
\end{figure}

The development of the magnetic order parameter was followed by single-crystal neutron diffraction measurements.
Rocking scans and integrated intensities of the $(001)$ reflection, with purely magnetic contribution, are shown in \fig\ref{fig:heidi}(a).
The $(001)$ reflection is split, as a consequence of the misalignment between the twins indexed 3 and 4.
The integrated intensity exhibits a sharp drop for increasing temperature.
It follows closely the order parameter modelled by the function $I\propto (1-T/T_N)^{\beta}$, as indicated by the solid line in \fig\ref{fig:heidi}(b).
The fit of the order parameter to the intensity of the $(001)$ reflection, as well as five other magnetic reflections (not shown), leads to a transition temperature $T_\mathrm{N}$=5.289\,(51)\,K and $\beta$=0.488\,(44) for the critical exponent.

The value determined here corresponds to the sharp peak in the specific heat at $T_\mathrm{N_2}$=5.4\,K, while there is no evidence of any additional feature coinciding with the broad shoulder at $T_\mathrm{N_1}$=5.6\,K.
However, the neutron diffraction data were recorded with a temperature step of $\approx$0.2\,K, which might be too coarse.

The determination of the magnetic order is based on the sets of reflections measured at 10\,K and 2.4\,K.
First, atomic positions were refined with the 10\,K dataset, with the same results as for the x-ray diffraction measurements listed in \tab\ref{tab:crystal-structure}.
These results were then used as fixed parameters for the refinement of the magnetic structure with the reflections measured at 2.4\,K.

\begin{figure}[t]
	\begin{center}
		\includegraphics[width=0.48\textwidth]{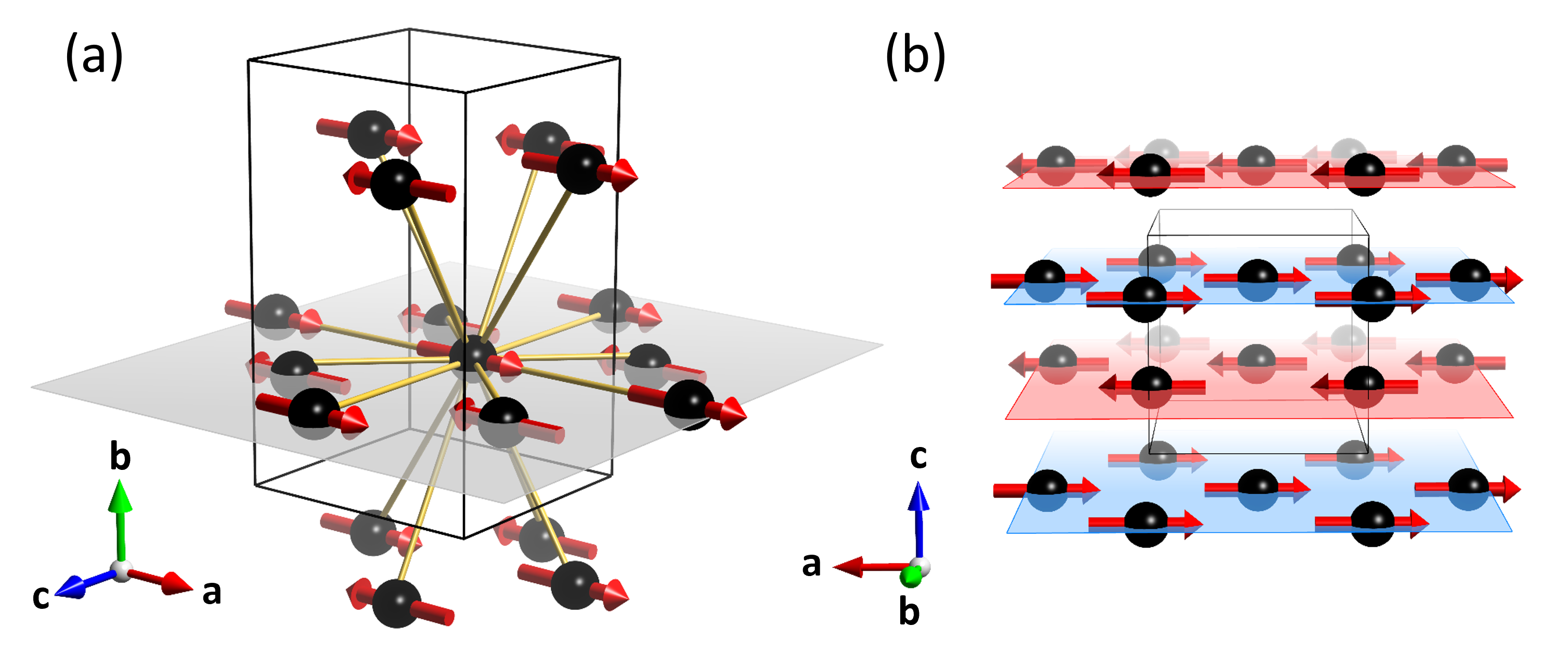}
	\end{center}
	\caption{Magnetic ordering in CePdAl$_3$. Magnetic moments are oriented along the $a$ axis in a collinear, antiferromagnetic arrangement. (a) Neighbors of the central cerium atom in the unit cell. The highlighted plane marks the distorted square-plane, where four nearest neighbours are aligned in an antiferromagnetic configuration with the central atom and four next-nearest neighbours in a ferromagnetic configuration. In addition, there are eight next-nearest neighbours out-of-plane, aligned in either antiferromagnetic or ferromagnetic configuration. (b) Magnetic moments within $ab$ planes (highlighted) form ferromagnetic sheets stacked antiferromagnetically along the $c$ axis.}
	\label{fig:magnetic-structure}
\end{figure}

The best refinement was obtained with the model described by the $Cmcm'$ magnetic space group (no. 63.461).
It describes collinear, antiferromagnetic order of magnetic moments located at the Ce sites, oriented along the $a$-axis, shown in \fig\ref{fig:magnetic-structure}. 
This corresponds to ferromagnetic order of the $ab$-planes, stacked along the $c$-axis, as shown in \fig\ref{fig:magnetic-structure}(c).
The refined value of the ordered magnetic moment is $\mu_{Ce}$=1.672\,(72)\,$\mu_B$.

\subsection{Crystal electric fields}
\label{sec:cef}

The $4f$ multiplet of the cerium atom with total angular momentum $J=5/2$ consists of six states.
In the orthorhombic CePdAl$_3$, the cerium atom is subject to crystal-electric fields with $m2m$ point group symmetry, which lifts the six-fold degeneracy into three Kramers' doublets, and allows for two CEF excitations from the ground state.
The crystal electric field is characterized by five $B_n^m$ parameters, with eigenvectors and excitation energies determined from \eq{\ref{eq:hamiltonian}}.

To investigate the crystal-electric fields in CePdAl$_3$ we performed inelastic neutron scattering (INS) measurements.
The measured crystal was oriented with the $(hkh)$ reflections in the scattering plane, a configuration approximately shared among all twin domains, with a mismatch of 3.7\degree.
Measurements of the crystal in this orientation allow to determine the spectral weight $S(\vec{Q}, E)$, for any vector $\vec{Q}$ with ($hkh$) crystal coordinates.

\begin{figure*}[t]
	\begin{center}
		\includegraphics[width=0.98\textwidth]{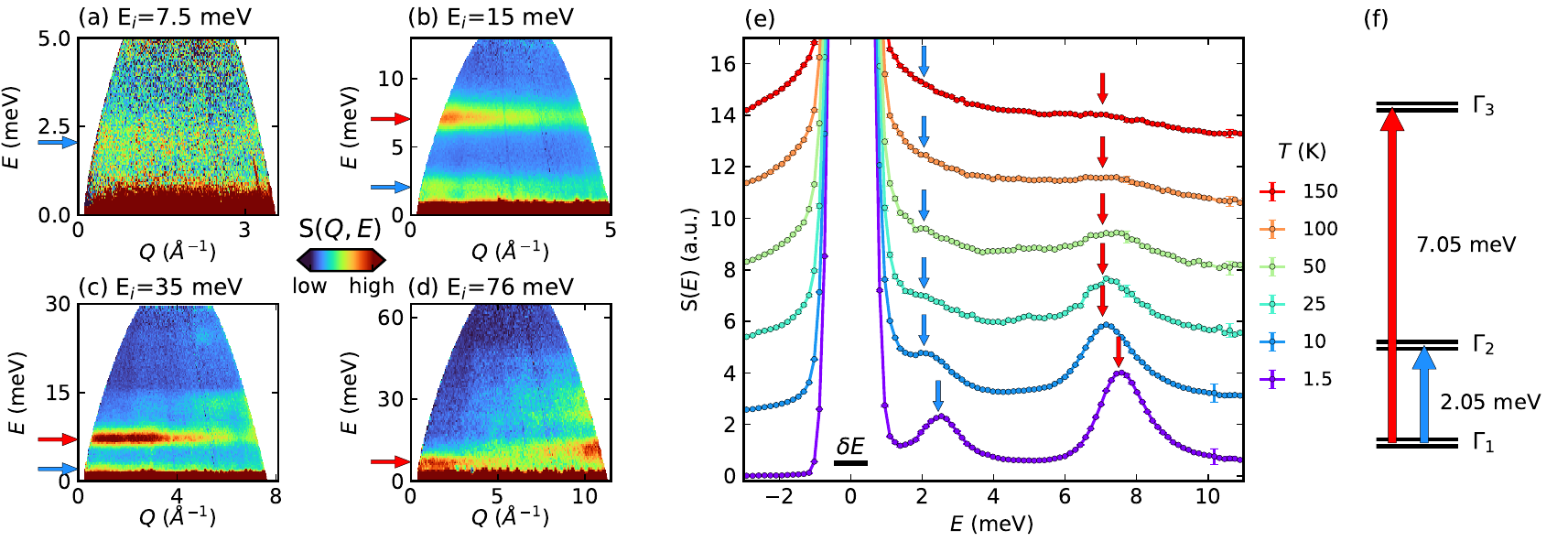}
	\end{center}
	\caption{Inelastic neutron scattering spectra of CePdAl$_3$. (a)-(d) Spectral weight $S(Q,\omega)$ measured at 10\,K with various incoming energies: (a) 7.5\,meV, (b) 15\,meV, (c) 35\,meV, and (d) 76\,meV. Two  bands of magnetic excitations are identified at positions marked with arrows. (e) Momentum integrated spectral weight $S(\omega)$ measured at various temperatures. Arrows indicate the position of fitted peak centres. Bar around zero corresponds to the instrumental energy resolution of $\delta E=$0.8 meV as determined from the width of the elastic line at 1.5\,K. (f) Sketch of the CEF scheme derived from the measurements and model.}
	\label{fig:cef-main}
\end{figure*}

The momentum-averaged ($|\vec{Q}|=Q$) and energy resolved spectral weight $S(Q, E)$ measured at 10\,K is depicted in \fig\ref{fig:cef-main}, panels (a-d).
In the low momentum transfers range, $Q<6$\,\AA$^{-1}$, the spectrum exhibits two dispersionless bands of excitations, characterized by a gradual decrease in spectral weight with increasing $Q$, indicating a magnetic character.
As the measurements were performed in the paramagnetic state, well above the magnetic ordering temperature $T_\mathrm{N}$, we interpret them as CEF excitations.

The determination of CEF transition energies is facilitated by examining the momentum-integrated spectral weight, $S(E)$, shown in \fig\ref{fig:cef-main}(e). 
In this case, the momentum $Q$ was integrated within the range $[0.8, 2.2]$\,\AA$^{-1}$ with the dataset measured at 10\,K and $E_i$=15\,meV as in \fig\ref{fig:cef-main}(b). 
The excitation energies are found to be $E_1$=2.05\,(10)\,meV and $E_2$=7.05\,(3)\,meV at 10\,K and at higher temperatures.
In the magnetically ordered state at 1.5\,K they shift to $E_1$=2.41\,(2)\,meV and $E_2$=7.44\,(2)\,meV, potentially due to internal fields.

\begin{figure}[t]
	\begin{center}
		\includegraphics[width=0.48\textwidth]{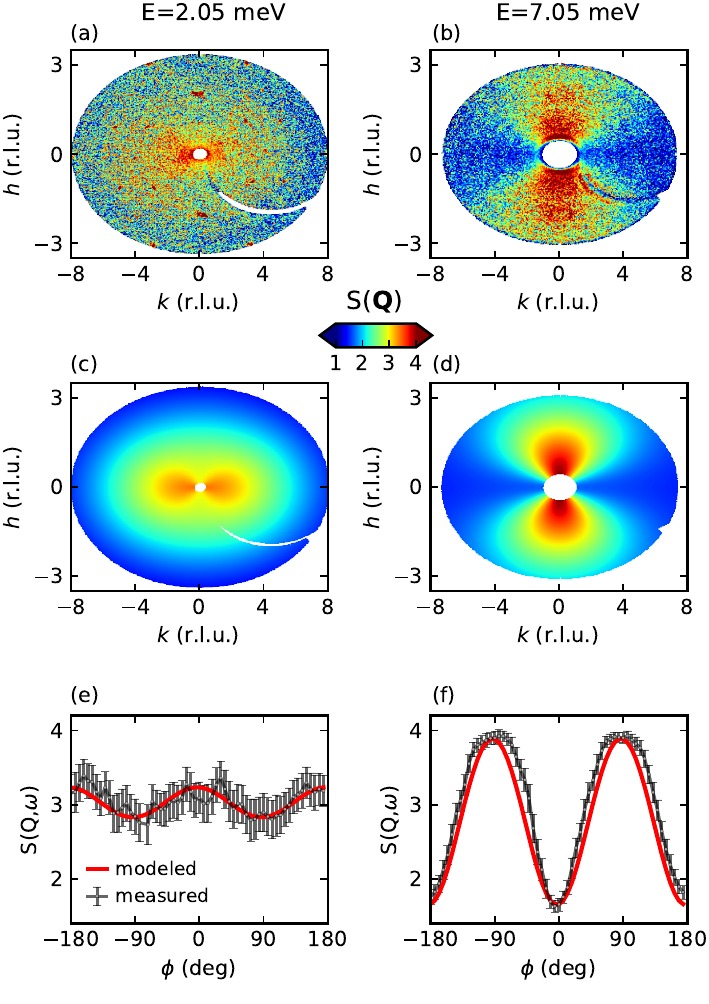}
	\end{center}
	\caption{Directional dependence of the spectral weight of CEF levels at 10\,K. Left column corresponds to the 2.05\,meV excitation, right column to the 7.05\,meV excitation. (a,b) Reconstruction of the measured spectral weight in the $(hkh)$ plane, obtained by integrating the energy in the vicinity of each level, as described in the main text. (c,d) Simulated spectral weight, based on parameters from \tab\ref{tab:CEF}. (e,f) Radially integrated intensity, comparison between measurement (black points) and simulations (red line). The arc with reduced intensity in the lower-right quadrant of measured maps is an artefact of the data collection.}
	\label{fig:cef-maps}
\end{figure}

In search for additional constraints on the $B_n^m$ parameters, we have investigated the angular dependence of the spectral weight at the energy transfer in vicinity of the excitation energy.
The characteristic distribution of intensity follows from the expectation value of the $\vec{\hat{J}}_{\perp Q}$ operator, as in \eq{\ref{eq:Jperp-SC}}.
The energy-integrated spectral weight obtained from measurements at 10\,K is shown in \fig\ref{fig:cef-maps}(a) and \fig\ref{fig:cef-maps}(b).
These were obtained by integrating $S(\vec{Q}, E)$ in the ranges $[1.5, 3.5]$\,meV and $[6, 9]$\,meV, respectively, to capture each transition.
Additionally, the angular distribution of the spectral weight for each energy level is displayed in \fig\ref{fig:cef-maps}(e) and \fig\ref{fig:cef-maps}(f).
The maps from \fig\ref{fig:cef-maps}(a-b) were radially integrated in the low-$\vec{Q}$ range $[0.75, 1.05]$\,\AA$^{-1}$, $\phi=0$ corresponds to the $\vec{Q}$=(0k0) direction and $\phi=90$\degree\ to $\vec{Q}$=(h0h).

To determine the Stevens parameters $B_n^m$, we followed the standard procedure employed in the analysis of CEF excitations.
In addition, we incorporated constraints derived from the angular distribution of spectral weight for each energy level.
The coordinate system for CEF calculation was taken with quantization axis $z$ along the $a$ crystal axis, $y$ along the $b$ axis and $z$ along the $c$ axis.
We performed a Monte Carlo search of $B_n^m$ parameters, imposing loose constraints on the energy and intensity distribution of the CEF levels.
Sets of parameters meeting these constraints were then refined against measured values, with an additional constraint on magnetization such that $M_{c^*} > M_{a^*} >M_{b^*}$ at 12\,T, as determined by magnetization measurements \cite{Kumar-PRRes-2023}.
Uncertainties of the $B_n^m$ parameters were estimated from the covariance matrix constructed during the refinement \cite{Press-2007}.

We tested $10^7$ sets of $B_n^m$ parameters, out of which $5.4 \cdot 10^4$ match the measured transition energies and only 826 further match the intensity distribution constraints.
For the 826 candidates, refinement converges into two unique sets of $B_n^m$ parameters, which are related to each other by rotating the reference frame by 90\degree\ around the $y$ axis, \ie, changing the signs of $B_2^0$ and $B_2^2$ parameters.
Such rotation corresponds to exchanging the $a$ and $c$ crystal axes, and is consistent with the ambiguity posed by measuring a twinned sample in the $(hkh)$ scattering plane.
Finally, we evaluate the ground state magnetic moment, which is $\pm$1.24\,$\mu_\mathrm{B}$ for the selected set and $\pm$0.47\,$\mu_\mathrm{B}$ for the discarded one.

The final set of $B_n^m$ parameters is presented in \tab\ref{tab:CEF}, along with eigenvectors and energy levels characterizing the crystal field in CePdAl$_3$.
Remarkably, the energies of CEF transitions align closely with the observed values, as well as the angular distribution of intensity, shown in \fig\ref{fig:cef-maps}.

The eigenvectors show that the ground state is dominated by the $\ket{\pm \frac{3}{2}}$ state, the first excited state by $\ket{\pm \frac{1}{2}}$, and the second excited state by $\ket{\pm \frac{5}{2}}$.
The expectation value of the magnetic moment for the ground state is 1.24\,$\mu_\mathrm{B}$ along the quantization axis $a$, close to the value determined from neutron diffraction 1.67\,$\mu_\mathrm{B}$.

Finally, we have not observed features characteristic of phonon-CEF coupling \cite{Loewenhaupt-PRL-1979, Adroja-PRL-2012, Cermak-PNAS-2019}, such as splitting of the CEF transitions in the $S(E)$ reconstruction shown in \fig\ref{fig:cef-main}, or anticrossing causing loss of intensity in $S(\vec{Q})$ at the energy of the CEF transition shown in \fig\ref{fig:cef-maps}.

%
%
%
\begin{table}[t]
	\centering
	\caption{Parameters and eigenvectors of the cerium $4f$ multiplet in the crystal electric field of orthorhombic CePdAl$_3$. Each eigenvector is double-degenerated, $\Gamma^{+} = \alpha \ket{\frac{5}{2}} + \beta \ket{\frac{3}{2}} + \gamma \ket{-\frac{1}{2}}$ and $\Gamma^{-} = \alpha \ket{-\frac{5}{2}} + \beta \ket{-\frac{3}{2}} + \gamma \ket{\frac{1}{2}}$. Entries list coefficients next to respective ket for either state.}
	\label{tab:CEF}
	\begin{tabularx}{0.48\textwidth}{X X X X X}
		\hline\hline
		\multicolumn{5}{l}{Crystal field parameters (meV)}\\
		\hline
		$B_2^0$ & $B_2^2$ & $B_4^0$ & $B_4^2$ & $B_4^4$ \\
		0.224 (16) & -0.200 (27) & 0.01408 (44) & -0.0458 (46) & 0.015 (10) \\
		\hline
		&&&&\\
		\multicolumn{5}{l}{Eigenvectors}\\
		\hline
		 & $\ket{\pm \frac{5}{2}}$ & $\ket{\pm \frac{1}{2}}$ & $\ket{\mp \frac{3}{2}}$ & E (meV)\\
		 $\Gamma_1^{\pm}$ & 0.096 & 0.105 &-0.990 & 0 \\
		 $\Gamma_2^{\pm}$ & 0.415 & 0.900 & 0.136 & 2.05 \\
		 $\Gamma_3^{\pm}$ & 0.905 &-0.424 & 0.043 & 7.05 \\
		\hline\hline
	\end{tabularx}	
\end{table}


\section{Discussion}

The combined x-ray and neutron diffraction measurements show that the CePdAl$_3$ sample investigated in this study crystallizes in the $Cmcm$ space group, with orthorhombic lattice.
It exhibits a pseudo-tetragonal twinning, suggestive of a high-temperature tetragonal to orthorhombic transition.

For further insights, we examine the synthesis protocols of CePdAl$_3$ from \reftaken\cite{Kumar-PRRes-2023, Franz-JAC-2016, Schank-JAC-1994}.
In particular, \citet{Schank-JAC-1994} report the synthesis of tetragonal CePdAl$_3$ with complete Pd-Al antisite disorder, which after annealing between 700-900\,\degree C transformed into an unknown phase.
The specific heat of this unknown phase shows the same signatures as the orthorhombic CePdAl$_3$ reported by \citet{Kumar-PRRes-2023} \ie, a sharp lambda peak with a shoulder on the high temperature side. 
Possibly, after annealing they obtained the orthorhombic phase of CePdAl$_3$, while their initial synthesis attempt, as well as that of \citet{Franz-JAC-2016}, resulted in quenching the high-temperature tetragonal phase, which is metastable at ambient conditions.
This scenario suggests that the orthorhombic CePdAl$_3$ is thermodynamically stable at ambient conditions, and that it undergoes a structural phase transition from tetragonal to orthorhombic structure between ambient temperature and 1000\,K, responsible for the pseudo-tetragonal twinning.

Our DFT-based calculations of Gibbs energy show that the tetragonal-orthorhombic transition takes place at 297\,K.
However, the calculations were performed on stoichiometric compounds and without atomic disorder, therefore the calculations might not accurately depict the thermodynamics of laboratory grown samples.
There seems to be a general tendency of the tetragonal Ce$T$Al$_3$ compounds to exhibit occupational disorder between $T$-Al sites \cite{Chlan-JPCM-2019, Klicpera-PRB-2015, Stekiel-PRR-2023}, which might also exist in the tetragonal CePdAl$_3$.

In orthorhombic CePdAl$_3$, the crystal-field splits the $4f$ multiplet into three Kramers doublets, as characterized in \tab\ref{tab:CEF}, with the values closely matching the results of inelastic neutron spectroscopy.
The orthorhombic distortion strongly affects the crystal field in CePdAl$_3$, as the constraints of the $4mm$ symmetry exhibited by Ce in tetragonal Ce$T$Al$_3$ systems permit only pure $\ket{\pm\frac{1}{2}}$ states, or mixing between $\ket{\pm\frac{3}{2}}$ and $\ket{\pm\frac{5}{2}}$ states. As reported in \tab\ref{tab:CEF}, we observe strong mixing between all base states. In addition, the values of $B_2^2$ and $B_4^2$ parameters are zero for $4mm$ symmetry, while we observe relatively large values of those parameters.

The full characterization of the CEF Hamiltonian allows to derive further observables and compare to the results of \citet{Kumar-PRRes-2023}.
Their analysis of the specific heat shows two CEF contributions at energies of 2.33\,(14) and 7.39\,(84)\,meV, consistent with our results.
Magnetization measured at 2.4\,K shows a kink around 5.5\,T that disappears above $T_\mathrm{N}$=5.3\,K.
The Zeeman energy at 5.5\,T is 0.7\,meV for the cerium moment, which is not enough to shift the next CEF level (2.05\,meV) to the ground state, and interpret the kink as a magnetic transition originating from CEF effects solely.
Thus, exchange interactions must play a significant role in this field-driven transition.
Magnetization saturation values measured at 14\,T and 10\,K, \ie, at temperature above long-range order where exchange interaction should play smaller role, are 1.13, 0.75, and 0.43\,$\mu_\mathrm{B}$ along the main axes, with corresponding calculated values of 1.13, 1.08, and 0.81\,$\mu_\mathrm{B}$, respectively, providing a good agreement for the model omitting exchange interactions.

The collinear arrangement of the magnetic moments in CePdAl$_3$, as well as the orthorhombic crystal structure, are reminiscent of CeAgAl$_3$, which exhibits ferromagnetic ordering below $T_\mathrm{C}$=3.8\,K, and also crystallizes in the $Cmcm$ structure \cite{Franz-JAC-2016, Nallamuthu-PhysicaB-2017}.
The level of distortion in CeAgAl$_3$ quantified from lattice parameters would correspond to a mismatch angle of 0.8\degree, significantly smaller than 3.7\degree\ for CePdAl$_3$.

An interesting observation is that orthorhombic Ce-113 compounds, CePdAl$_3$ and CeAgAl$_3$, establish collinear magnetic order with zero modulation vector, while \textit{all} of the reported tetragonal systems establish non-zero modulation.
The modulation is either commensurate, when magnetic moments are aligned with the four-fold axis as in Ce$T$Ge$_3$ with $T$=Rh, Ir, Co \cite{Hillier-PRB-2012,Anand-PRB-2018,Smidman-PRB-2013}, or incommensurate when moments are aligned in-plane ($m_z$=0), as in CeAuAl$_3$ \cite{Adroja-PRB-2015}, CePtAl$_3$ \cite{Stekiel-PRR-2023}, and CeCuGa$_3$ \cite{Anand-PRB-2021}.
An exception from the latter case seems to be posed by CeCuAl$_3$ \cite{Klicpera-PRB-2015}, which establishes in-plane magnetic order with commensurate modulation of five unit cells.

This observation leads to the hypothesis that the orthorhombic deformation is at the origin of the non-modulated magnetic order.
To explore that hypothesis we examine magnetic interactions in CePdAl$_3$, assuming they follow the RKKY interaction model \cite{vVleck-RevModPhys-1962}, typical for metallic systems with localised magnetic moments.
Within the RKKY model the strength and the sign of exchange interactions are determined solely by the distance between magnetic atoms, Ce-Ce pairs in this case.
As described in \fig\ref{fig:magnetic-structure}(a) there are (i) four in-plane nearest neighbours to the central ion, (ii) additional four in-plane next-nearest neighbours, and (iii) eight out-of-plane next-nearest neighbours.

Considering tetragonal CePdAl$_3$ with lattice parameters $a_t$=4.37\,\AA\ and $c_t$=$b$=10.4\,\AA, the four-fold symmetry restricts each of (i), (ii), and (iii) as equidistant classes of neighbours with (i) at 4.37\,\AA, (ii) at 6.18\,\AA, and (iii) at 6.05\,\AA.
With assignment of different coupling to each class, (i) $J_x^\mathrm{NN}$, (ii)  $J_x^\mathrm{NNN}$, and (iii) $J_z^\mathrm{NNN}$, we examine the case of antiferromagnetic nearest neighbour interaction $J_x^\mathrm{NN}$ as a leading term in the interaction scheme.
This causes the in-plane nearest neighbours to align antiferromagnetically with respect to the central atom, but also the four out-of-plane atoms with respect to each other, as in \fig\ref{fig:magnetic-structure}(a).
Then, regardless of the sign of $J_z^\mathrm{NNN}$, the interaction between the central atom and the out-of-plane atoms is frustrated, as two pairs are aligned ferromagnetically and two others antiferromagnetically.
Also, if we assume the $J_z^\mathrm{NNN}$ to be the leading term, by similar line of reasoning we arrive at conclusion that antiferromagnetic $J_x^\mathrm{NN}$ interaction frustrates the nearest neighbour interactions.
The in-plane, next-nearest neighbour interaction $J_x^\mathrm{NNN}$ of ferromagnetic character is compatible with antiferromagnetic $J_x^\mathrm{NN}$, but antiferromagnetic coupling again frustrates the system, as it competes with the nearest neighbour interaction $J_x^\mathrm{NN}$.
Most importantly, the frustration arises independently on the orientation of the magnetic moments.

According to the RKKY model, the exchange interaction is an oscillating function of atomic distance.
This allows for significant next-nearest neighbour interaction strength, such that the frustration contributes on a significant scale to the overall energy of the system, and does not permit a collinear ground state.

Orthorhombic distortion in CePdAl$_3$ preserves the nearest neighbour bonds, while splitting and mixing next-nearest neighbours, such that (ii) and (iii) fall within a range from 5.72\,\AA\ to 6.38\,\AA\ \footnote{Orthorhombic distortion splits the 4$\times$6.18\,\AA\ in-plane, next-nearest neighbor (NNN) into 2$\times$5.97\,\AA\,(FM) and 2$\times$6.38\,\AA\,(FM) neighbors, while 8$\times$6.05\,\AA\ out-of-plane NNN into 2$\times$5.72\,\AA\,(AF), 4$\times$6.1\,\AA\,(FM) and 2$\times$6.28\,\AA\,(AF).}.
Consequently, the frustration imposed by equivalence of out-of-plane bonds is relieved in the orthorhombic structure.
This allows to realize the arrangement exhibited by CePdAl$_3$ shown in \fig\ref{fig:magnetic-structure}(a).

Among tetragonal Ce-113 compounds, CeCuAl$_3$, CePtAl$_3$, and CeCuGa$_3$ seem to realize the antiferromagnetic nearest neighbour interaction, based on their magnetic moments arrangement.
In the magnetically ordered state they establish modulation of the magnetic moment amplitude with an in-plane modulation vector, $k_z$=0, which breaks the four-fold symmetry axis and lowers the lattice symmetry from tetragonal to orthorhombic.
The modulation has the amplitude of $\mu_\mathrm{max}$=0.2\,$\mu_\mathrm{B}$ in CeCuAl$_3$, 1.1\,$\mu_\mathrm{B}$ in CePtAl$_3$, and 0.67\,$\mu_\mathrm{B}$ in CeCuGa$_3$.
As the crystal field reduces the magnetic moment in CePdAl$_3$ only to 1.67\,$\mu_\mathrm{B}$ from the expected $g_\mathrm{J}J$=2.14\,$\mu_\mathrm{B}$ for the isolated ion, further reduction of the magnetic moment amplitude in CeCuAl$_3$, CePtAl$_3$, and CeCuGa$_3$ could originate from the frustration.

In summary, the restrictions of tetragonal symmetry frustrate the magnetic interaction scheme in CePdAl$_3$ and do not allow for the long-range magnetic order.
The orthorhombic distortion relieves the frustration and allows for a long-range antiferromagnetic state.


\section{Conclusions}

We investigated the crystal and magnetic structure of orthorhombic CePdAl$_3$ by x-ray and neutron diffraction, with further insights into crystal-field effects based on inelastic neutron scattering.

The systematic pseudo-tetragonal twinning of the orthorhombic lattice of CePdAl$_3$ originates from lattice deformation following the tetragonal-orthorhombic transition.
The transition has its grounds in the small energy differences between tetragonal and orthorhombic phases, which allow quenching either phase, consistent with previous reports.

Orthorhombic CePdAl$_3$ establishes a collinear, antiferromagnetic type of long-range magnetic order below $T_\mathrm{N}$=5.3\,K.
The magnetic moments of the cerium atoms order along the $a$ axis and form ferromagnetic sheets, stacked along the $c$ axis.
This type of order is consistent with single ion anisotropies in form of $B_n^m$ Stevens parameters, which arise from crystal-electric field effects.
The CEF scheme is fully characterized and the measured spectra are in perfect agreement with our calculations based on the determined $B_n^m$ parameters.

We discussed the frustration mechanism based on the antiferromagnetic interaction between nearest neighbours on the tetragonal lattice, and how the orthorhombic distortion releases that frustration.
We propose that this mechanism is responsible for establishing long-range magnetic order in CePdAl$_3$, as opposed to tetragonal CePdAl$_3$, where no order was observed down to 0.1\,K.


\section{Acknowledgements}

We kindly acknowledge Thomas M\"{u}ller for help with the analysis of the DNS data, and Anatoliy Senyshyn for help in the analysis of diffraction data.
We also thank the staff at the MLZ and ILL for support.
Parts of the neutron diffraction measurements were performed on HEiDi, jointly operated by the RWTH Aachen University and JCNS within the JARA-FIT collaboration.

This study was funded by the Deutsche Forschungsgemeinschaft under project No.323760292 (Mehrkomponentige Elektronische Korrelationen in Nicht-Zentrosymmetrischen f-Elektron-Verbindungen), 
TRR80 (From Electronic Correlations to Functionality, Project No. 107745057), TRR360 (Constrained Quantum Matter, Project No. 492547816), SPP2137 (Skyrmionics, Project No. 403191981, Grant PF393/19), and the excellence cluster MCQST under Germany's Excellence Strategy EXC-2111 (Project No. 390814868). Financial support by the European Research Council (ERC) through Advanced Grants No. 291079 (TOPFIT) and No. 788031 (ExQuiSid), the Czech Science Foundation GA\v{C}R under the Junior Star grant No. 21-24965M (MaMBA) for P.C and No. 22-35410K for D.L.,  No. CZ.02.01.01/00/22\_008/0004572 and e-INFRA CZ (ID:90254) by MEYS of Czech Republic are gratefully acknowledged.


%
%



\end{document}